\documentclass{pazhastl}
\usepackage{graphicx}
\usepackage{natbib}

\begin{document}

\journalinfo{2007}{33}{12}{804}[806]

\title{On the Variability of Gamma-Ray Burst Afterglows --- A~Possibility of
  a Transition to Nonrelativistic Motion}

\author{R.~A.~Burenin\address{1}\email{rodion@hea.iki.rssi.ru}
  \addresstext{1}{Space Research Institute (IKI), ul. Profsoyuznaya 84/32,
    Moscow, Russia} }

\shortauthor{R.~A.~Burenin}  
\shorttitle{On the variability of GRB afterglows}
\submitted{June 13, 2007}

\begin{abstract}
  
  Variability on time scales $\delta t<t$ is observed in many gamma-ray
  burst afterglows. It is well known that there should be no such
  variability if the afterglow is emitted by external shock, which is
  produced by the interaction of ultrarelativistic ejecta with the ambient
  interstellar medium, within the framework of simple models. The
  corresponding constraints were established by \cite{ioka} and in some
  cases are inconsistent with observations. On the other hand, if the motion
  is not relativistic, then the fast variability of the afterglow can be
  explained much more easily.

  In this connection we discuss various estimates of the time of the
  transition to subrelativistic motion in GRB source. We point out, that
  this transition should occur on an observed time scale of $\sim10$ days.
  In the case of a higher density of the ambient interstellar medium
  $\sim10^{2}$--$10^{4}$~cm$^{-3}$ or dense stellar wind with $\dot M \sim
  10^{-5}$--$10^{-4}~M_\odot$/year the transition to a subrelativistic
  motion can occur on a time scale of $\sim1$~day. These densities may well
  be expected in star-forming regions and around massive Wolf-Rayet stars.

\end{abstract}

\section*{Introduction}
\label{sec:intro}

According to current models of gamma-ray bursts (GRBs), ultrarelativistic
motion of the jet pointing toward the observer takes place in their sources
\citep[see reviews by][]{meszaros,zhang,piran}. Their afterglows in X-ray,
optical, and radio bands are explained by the radiation that emerges at the
front of the external shock produced by jet interaction with the ambient
interstellar medium surrounding the source. In this case, a variable light
curve cannot emerge from a homogeneous spherical emitting shell, that moves
inside a cone with the opening angle $\theta > \gamma^{-1}$, where $\gamma$
--- the Lorentz factor of the jet. For the remote observer, the variability
time scale should be $\delta t \sim t$.

However, the variability of optical afterglows at time scales $\delta t < t$
is observed in many cases. The best-known is GRB 030329 whose afterglow was
studied in great detail due to its exceptional brightness
\citep{burenin,urata,lipkin}. Similar variability was also observed in other
cases where detailed measurements of the light curve were obtained: GRB
021004 \citep[see e.g., ][]{holland,bersier,antonio05}, 050408
\citep{antonio07}, 060526 \citep{dai,khamitov2} and others.

Various explanations for the variability of GRB afterglows were discussed in
detail by \cite{ioka}, who also established constraints on the afterglow
variability time scales and amplitudes for various models. They are all
based on the assumption of ultrarelativistic motion of the jet.  In some
observed afterglows, the constraints obtained in simple models of the jet
are violated \citep{ioka,khamitov2}.

However, the motion may well become moderately relativistic on the observed
time scales in case of higher density of the ambient medium or in presence
of a dense stellar wind. For example, \cite{dailu99} discussed this
possibility for the afterglow of GRB 990123 and showed that the transition
to a nonrelativistic motion on a time scale of about 2.5 days occurs if the
density of the surrounding medium is $\sim\!\!10^6$~cm$^{-3}$. In this
paper, the sideways expansion of the jet once the condition $\gamma <
\theta^{-1}$ is satisfied \citep{rhoads,rhoads2,sari99} was not taken into
account. Because of this expansion, the jet collects more material on its
way and slows down faster.

In this note, we discuss various estimates of the time of the transition to
subrelativistic motion and show that it may well be $\sim1$ day.

\section{When does the ultrarelativistic motion~end?}
\label{sec:nr}

The most simple estimate of the time pf the transition to nonrelativistic
motion, $t_{NR}$, taking into account the sideways expansion of the jet, was
obtained by \cite{waxman98}:
\begin{equation}
  \label{eq:waxman}
  t_{NR}\approx 27 (E_{52}/n_1)^{1/3}\theta_{0.1}^{2/3}~\mbox{days},
\end{equation}
where $E_{52}$ is the ``isotropic'' energy of the shell in units of
$10^{52}$~erg/s/cm$^2$, $n_1$ is the particle number density in the medium
surrounding the source in units of cm$^{-3}$, and $\theta_{0.1}=\theta/0.1$
--- is the jet opening angle.

This estimate was obtained using the dependency $\gamma(t)$ for adiabatic
shell and assuming that the motion after the onset of jet sideways expansion
approaches a spherically symmetric one and take place according to the same
adiabatic solutions. In order of magnitude, it agrees well with the other
estimates that can be obtained from a more comprehensive analysis of the jet
dynamics. For example, if the jet begins to expand at $\gamma=\theta^{-1}$,
then, given that after that the gamma factor depends on the observed time as
$\gamma\propto t^{-1/2}$ \citep{rhoads2}, we will obtain:
\begin{equation}
  t_{NR} = t_j \theta^{-2}~,
\end{equation}  
where $t_j$ is the time of the beginning of jet sideways expansion. If we
take the expression for $t_j$ from \cite{sari99}, then $t_{NR} = 26$~days
with the same parameter dependency as above.

\cite{rhoads2} assumed that the jet in its rest frame expands with the speed
of sound $c_s=c/\sqrt{3}$, not with the speed of light. Accordingly, the jet
expansion begins later, at $\gamma\approx (3\sqrt{2}\theta)^{-1}$. Using the
$\gamma(t)$ dependency from this work and applying the same inference as
above, we obtain $t_{NR} = 25$~days for the same parameters. \cite{panmes}
assumed that the jet also expands with the speed of sound.  They pointed out
that, when $\theta > 0.1$, the shell ceases to be ultrarelativistic even
before the onset of the sideways expansion. In this work, $t_{NR}$ was
estimated explicitly. Rewriting it with our parameters, we obtain $t_{NR} =
6.1$~days, with the same parameter dependency as in (\ref{eq:waxman}).

If we take into account that initially the motion of the shell may not be
adiabatic and it can lose a substantial part of its energy through
radiation, then the estimate of $t_{NR}$ can decrease significantly. For
synchrotron radiation, the motion should be radiative if much of the energy
that the material gains at the shock front goes into the accelerated
electrons and if the electrons cool down rapidly compared to the dynamical
time. The latter always holds at the beginning of shell motion. The
expression for the energy, to be used to calculate the subsequent adiabatic
motion if the evolution was initially radiative is given by \cite{sari98}.
Substituting this energy into (\ref{eq:waxman}) yields:
\begin{equation}
  \label{eq:rad}
  t_{NR,r}\approx 7.6~\epsilon_B^{-1/5}\epsilon_e^{-1/5}
  E_{52}^{4/15} \gamma_2^{-4/15} n_1^{-7/15} \theta_{0.1}^{2/3}~\mbox{days},
\end{equation}
where $\epsilon_B$ and $\epsilon_e$ are the energy density fractions of the
magnetic field and accelerated electrons behind the shock, $\gamma_2 =
\gamma_0/100$ is the initial gamma factor of the ejecta. This estimate
should be used if it is assumed that $\epsilon_e\sim 1$, i.e., the electrons
are accelerated effectively at the shock front.

The estimate of $t_{NR}$ for a stellar wind with a density proportional to
$r^{-2}$ can be taken from \cite{livio}. Rewriting it with our typical
parameters, we will obtain:
\begin{equation}
  \label{eq:wind}
  t_{NR,w}\approx 5.7~E_{52} (\dot M_{-5}/v_{3})^{-1} 
  \theta_{0.1}^{2}~\mbox{days},
\end{equation}
where $\dot M_{-5}$ is the mass loss rate of the star in units of
$10^{-5}~M_\odot$year$^{-1}$ and $v_{3}$ is the wind velocity in units
$10^{3}$~km~s$^{-1}$. If the electrons are accelerated effectively and the
evolution is initially radiative, then this estimate should be much smaller,
just as in the above case of a constant-density medium.

\section{Discussion}
\label{sec:discuss}

From these estimates we see, that the observed transition time to
nonrelativistic motion is $\sim10$~days even for the commonly accepted
typical parameters. A higher density of the ambient medium $n\sim
10^{4}$~cm$^{-3}$ for an adiabatic shock, or $n\sim 10^{2}$~cm$^{-3}$, if
the electrons at the shock front are accelerated effectively and the shell
is initially radiative, or a stellar wind with $\dot M \sim
10^{-5}$--$10^{-4}~M_\odot$~year$^{-1}$ is required for $t_{NR}\sim1$~day.

A higher density of the interstellar medium is actually may well be expected
near GRB sources, since at least a large fraction of them are related to
supernova explosions at the end of the evolution of massive stars and occur
in regions of enhanced star formation \citep[see, e.g., the review by][and
references therein]{woosley}. In addition, massive stars at the end of their
evolution intensively lose their outer layers and should be surrounded by
dense stellar winds \citep[e.g.,][]{crowther}. In some cases, the presence
of a wind around GRB sources is confirmed observationally. For example,
high-resolution spectroscopy of the GRB 021004 afterglow shows that its
progenitor was a massive Wolf-Rayet star surrounded by a wind with a mass
loss rate of $\sim 10^{-4}~M_\odot$~year$^{-1}$ and an the expansion
velocity up to $3000$~km~s$^{-1}$ \citep{mirabal,lazzati}.

If the motion is no longer ultrarelativistic, then the fast variability of
the afterglow on a time scale $\delta t \ll t$ can be explained, for
example, by the presence of density inhomogeneities. In fact, the presence
of these inhomogeneities can also be expected. For example, the stellar
winds around Wolf-Rayet stars are known to be highly inhomogeneous and to
have a clumpy structure \citep{hamann,crowther}. In addition, the cooling
time in the observed optical part of the spectrum should be short.  This
holds for the synchrotron spectrum at late evolutionary phases of the shock
\citep[e.g.,][]{sari98}.

Thus, the transition to subrelativistic motion on the observed time scale of
$\sim1$~day is actually quite possible and, in this case, the fast
variability of GRB afterglows can be easily explained. Of course, the fast
variability of GRB afterglows can also be explained under special
assumptions in the case of an ultrarelativistic jet \citep{ioka}.  However,
the transition to a nonrelativistic motion should at least be considered as
one of the possible explanations of this fast variability.  This requires a
higher density interstellar medium or dense stellar wind, which are expected
in star forming regions and around massive Wolf-Rayet stars.

\acknowledgements

The author is grateful to the anonymous referee who made several important
remarks regarding the content of the paper. This work was supported by the
grants RFFI 05-02-16540, RFFI 07-02-01004, NSh-1100.2006.2, MK- 4064.2005.2,
and Programs P-04 and OFN-17 of the Russian Academy of Sciences.

\end{document}